\documentstyle[12pt,twoside]{article}
\def\eqn#1{(\ref{#1})}
\def\vec#1{{\bf #1}}
\newcommand{\dfourx}{d^4\hskip-.1em{x}}
\newcommand{\dfoury}{d^4\hskip-.1em{y}}
\newcommand{\dfour}{d^4\hskip-.1em}
\newcommand{\Pmn}{P^{\mu\nu}}
\newcommand{\rmn}{\rho^{\mu\nu}}
\newcommand{\Pimn}{\Pi^{\mu\nu}}
\newcommand{\Pimnr}{\Pi^{\mu\nu}_R}
\newcommand{\Pimnt}{\Pi^{\mu\nu}_T}
\newcommand{\Pimnxy}{\Pi^{\mu\nu}(x,y)}
\newcommand{\Pimnsub}{\Pi^{\mu\nu}_{\rm sub}}
\newcommand{\amx}{A_\mu(x)}
\newcommand{\amk}{A_\mu(k)}
\newcommand{\ank}{A_\nu(k)}
\newcommand{\amdk}{A_\mu^\dagger (k)}
\renewcommand{\Im}{{\rm\ Im\ }}
\renewcommand{\Re}{{\rm\ Re\ }}
\newcommand  {\tr}{{\rm\ tr\ }}
\newcommand{\lessim}
{\lower0.6ex\hbox{\vbox{\offinterlineskip\hbox{$<$}\vskip1pt\hbox{$\sim$}}}}

\newcounter{subseqn}
\newcounter{saveeqn}

\renewcommand{\thesection}%
 {\arabic{section}\setcounter{equation}{0}}
\renewcommand{\thesubsection}%
 {\setcounter{subseqn}{\value{equation}}%
\thesection\alph{subsection}%
\setcounter{equation}{\value{subseqn}}}
\renewcommand{\theequation}%
 {\mbox{\arabic{section}.\arabic{equation}}}

\newcommand{\alpheqn}%
 {\setcounter{saveeqn}{\value{equation}}%
\stepcounter{saveeqn}\setcounter{equation}{0}%
\renewcommand{\theequation}%
 {\mbox{\arabic{section}.\arabic{saveeqn}\alph{equation}}}}
\newcommand{\reseteqn}{\setcounter{equation}{\value{saveeqn}}%
\renewcommand{\theequation}{\mbox{\arabic{section}.\arabic{equation}}}}

\begin{document}
\thispagestyle{empty}
\begin{flushright}
CU-TP 594 \\
CTP\#2205 \\
hep-ph/9305241 \\
May 1993
\end{flushright}

\vspace{\fill}

\begin{center}
{\LARGE High Temperature Response Functions \\
\vspace{6pt}
and the Non-Abelian Kubo Formula*} \\
\vspace{\fill}
{\large R.~Jackiw} \\
\vspace{2pt}
{\em Center for Theoretical Physics \\
Laboratory for Nuclear Science \\
Massachusetts Institute of Technology \\
Cambridge, MA\ \ 02139} \\
\vspace{7pt}
{\large and \\
\vspace{9pt}
V.~P.~Nair} \\
\vspace{2pt}
{\em Physics Department \\
Columbia University \\
New York, NY 10027}
\end{center}

\vspace{\fill}
\begin{abstract}
We describe the relationship between time-ordered and retarded
response functions in a plasma.
We obtain an expression,
including the proper $i\epsilon$-prescription,
for the induced current due to hard thermal loops in a
non-Abelian theory,
thus giving the non-Abelian generalization of the Kubo formula.
The result is closely related to the eikonal for a Chern-Simons
theory and is relevant for a gauge-invariant description of
Landau damping in the quark-gluon plasma at high temperature.
\end{abstract}
\vspace{\fill}
\begin{center}
Submitted to {\it Physical Review D\/} 15
\end{center}
\vspace{\fill}
\hbox to \hsize{\hrulefill}
\noindent
* This research is supported in part by the U.~S.~Department of Energy.

\pagebreak

\thispagestyle{empty}
\begin{center}
{\bf High Temperature Response Functions
and the Non-Abelian Kubo Formula} \\
{\rm R.~Jackiw and V.~P.~Nair}
\end{center}

\section{Introduction}
The behavior of electromagnetic fields in a plasma of charged
particles is described by the polarization tensor
$\Pimnxy$, which is the two-point current correlation function;
perturbatively this is a one-charged-particle-loop  diagram with
two external photon  lines.  The real part of this tensor
describes phenomena such as Debye screening and propagation of
plasma waves; the imaginary part describes the damping of fields
in  the plasma (Landau damping) \cite{1}.  If we integrate out the
charged fields in a functional integral for the theory, the
polarization tensor naturally emerges as the thermal average of
the {\em time-ordered product} of two currents.  However, there
are situations where the response of the plasma
to the electromagnetic field is described as the average of the
{\em retarded commutator} of currents.   For the real part of the
response function, the retarded commutator and  the time-ordered
product coincide.  We calculate the imaginary part for both and
discuss their relationship in terms  of spectral representations \cite{2}.

In a  non-Abelian plasma,  such as the one arising perhaps from
quarks and gluons,
the response  of the plasma to fields is more complicated.  At high
temperature, a perturbative approach can be used.  Nevertheless,
it is necessary to go beyond the two-point function; one must
include at least part of the contribution from  all higher point
functions for reasons of gauge invariance.  There is also the need
for a resummation of perturbation  theory \cite{3}.  One must
first sum up `hard thermal loops' to  define new effective
propagators and  vertices.  Hard thermal  loops are thermal
one-loop diagrams in which the external momenta are relatively
soft ($\sim g T$,  where $g$ is the coupling constant and $T$ is
the temperature, taken to be high)
while  the loop momentum is relatively hard
($\sim T$).
One must reorganize perturbation theory in
terms of effective vertices and propagators defined
by these hard thermal loops,
before integration on small values
(\lessim\ $g T$)
of loop momenta
can be carried out \cite{4}.
This is necessary so  that
all contributions to  a given order in  the coupling constant can
be consistently included.

Recently it was shown \cite{5} that the
generating functional for hard thermal loops, or equivalently the
effective action which describes Debye screening, plasma waves,
{\it etc.}, is given in terms of the eikonal for a Chern-Simons theory,
which had been constructed earlier in a non-thermal context
\cite{6}.  We now describe how this result is extended
to include  some of the decay and damping effects.  Our discussion
is still at the level of hard  thermal loops; we do not address
effects that arise by incorporating effective propagators
and vertices in subsequent soft momentum integrations.  Our
result can be considered as a non-Abelian generalization of the
Kubo formula \cite{7}.

In  Section II we present  the time-ordered and retarded
polarizations for the Abelian plasma  at high temperature.  The
physical significance of the imaginary part is explained in
Section  III.  The induced current due to  hard thermal loops in a
non-Abelian theory is constructed
in Section IV.

\section{Response Functions for the Abelian Plasma}

We consider QED with the fermion current
$J^\mu = e \bar{\psi} \gamma^\mu \psi$
and interaction Lagrangian
${\cal L}_{\rm int} = - A_\mu J^\mu$.
The equations of motion for the gauge field $A_\mu$ have $J^\mu$
as a source  term.
The calculation of the current $J^\mu$ is thus a suitable way of
studying response functions.  The scattering operator $S$ is
given by $T \exp [ -i \int {\dfourx A_\mu j^\mu} ]$ where
$j^\mu = e {\bar{\psi}}_I \gamma^\mu \psi_I$
is the current in the interaction picture, denoted by the
subscript $I$.  We can write $J^\mu$ in  terms of $S$ as
\begin{equation}
J^\mu (x) = i S^{-1} \frac{\delta S}{\delta A_\mu (x)} \label{2.1}
\end{equation}
The equation  of motion for the field $A_\mu$ is
\begin{equation}
\partial_\nu F^{\nu\mu} (x)
= i S^{-1} \frac{\delta S}{\delta A_\mu (x)} \label{2.2}
\end{equation}
[While no gauge  choice is specified in \eqn{2.2}, it can be
fixed for example by adding $-\frac{1}{2} (\partial \cdot A)^2$
to the Lagrange density,
whereupon the left side would become $\Box A^\mu$ (Feynman
gauge).]
We can use \eqn{2.1} to obtain $J^\mu$ as a power series in
$A_\mu$.
In particular, a Taylor expansion of \eqn{2.1} to linear order
in $A$ gives
\begin{equation}
J^\mu (x) = j^\mu (x) - i \int
{\dfoury\ \ \theta (x^0 - y^0) [ j^\mu (x), j^\nu (y) ] A_\nu (y)}
\label{2.3}
\end{equation}
Using this in \eqn{2.2} and taking a  thermal average we can write
\begin{equation}
\partial_\nu  F^{\nu\mu} (x) = \int
{\dfoury \Pimnr (x, y) A_\nu (y)}
\label{2.4}
\end{equation}
where
\begin{equation}
\Pimnr (x, y) = -i \theta (x^0-y^0)
\ \langle [ j^\mu (x),  j^\nu (y) ] \rangle
\label{2.5}
\end{equation}
The angular brackets denote thermal averaging, with the unperturbed density
matrix $e^{-H_0 / T}$, so that $\langle j^\mu \rangle$ vanishes. (The
Boltzmann constant is set to unity.)  Thus the response function \eqn{2.5},
{\it viz.} the average of the retarded commutator [or equation \eqn{2.4}],
is appropriate to  the situation where we
perturb the plasma by the field and ask how the field evolves.
Equation \eqn{2.5} is the Kubo formula.

The effective action $\Gamma (A)$, which is the sum of fermion loops,
is given by
\begin{equation}
e^{i \Gamma (A)} = \langle T e^{-i\int{A_\mu j^\mu}} \rangle
\label{2.6}
\end{equation}
The current that is here naturally defined as
${\cal J}^\mu (x) = -\frac{\delta\Gamma (A)}{\delta A_\mu (x)}$
is given to  linear order in $A_\mu$ by
\begin{equation}
{\cal J}^\mu (x) = \int {\dfoury \, \Pimnt (x, y) A_\nu (y)}
\label{2.7}
\end{equation}
where
\begin{equation}
\Pimnt (x, y) = -i \langle T j^\mu (x) j^\nu (y) \rangle
\label{2.8}
\end{equation}
Clearly, the two response functions do not coincide.
One difference is immediately evident:  The time ordered product is symmetric
in its labels $(\mu, \nu) (x, y)$ [as it must be since
$\Pimnt$ arises from the generating functional
$\Gamma (A) \sim 1-\frac{1}{2}
\int  {\dfourx \dfoury A_\mu (x) \Pimnt (x, y) A_\nu (y) + \cdots}$ ]
while the retarded commutator does not enjoy any such symmetry.

To evaluate \eqn{2.5} and \eqn{2.8}, we  make use of the free-field
commutator.
\begin{equation}
\left\{ \psi (x), \bar{\psi} (y) \right\}
= \int {\frac{d^3 p}{(2\pi)^3}\,\frac{\gamma\cdot p}{p_0}\,\cos p(x-y) }
\label{2.9}
\end{equation}
($p_0 = \sqrt{\vec{p}^2}$;  since we are interested in the high temperature
contribution, fermion masses are negligible by comparison.)
The averages needed in the computation can be evaluated using the following
expression for the propagator.
\begin{eqnarray}
S (x, y) &=& \langle T \psi (x) \bar{\psi} (y) \rangle \nonumber\\
&=& \int { \frac{d^3 p}{(2\pi)^3}\ \frac{\gamma\cdot p}{2 p_0} }
\ \biggl\lbrace \theta (x^0 - y^0)
\left( \alpha_p e^{-ip(x-y)} + \bar{\beta}_p e^{ip(x-y)} \right) \\
&& \hskip.9in \mbox{} - \theta (y^0 - x^0)
\left( \beta_p e^{-ip(x-y)} + \bar{\alpha}_p e^{-ip(x-y)} \right)
\biggr\rbrace \nonumber
\label{2.10}
\end{eqnarray}
where
\[ \alpha_p = 1-n_p, \qquad \beta_p = n_p \]
\begin{equation}
n_p = \frac{1}{e^{p_0/T} + 1}
\label{2.11}
\end{equation}
(The bar refers to antifermion distributions.)
In evaluating the two-point functions,  we shall use \eqn{2.9}, \eqn{2.10}
and carry out the time-integrations first, introducing as needed
convergence factors
$e^{\pm\epsilon y^0}$, $\epsilon$ small and positive.  We then
get energy-denominators with $i\epsilon$-terms.  Further simplification and
extraction of the imaginary part can be easily done at this stage.  We find
for the polarization tensor,
\alpheqn
\begin{equation}
\Pimn (x,y) =
\int { \frac{\dfour{k}}{(2\pi)^4} e^{-ik(x-y)} \Pimn (k) }
\label{2.12a}
\end{equation}
\begin{eqnarray}
\Pimn (k) &=&
\int {\frac{d^3 q}{(2\pi)^3}\,\frac{1}{2 p_0}\,\frac{1}{2 q_0}
\biggl\lbrack
T^{\mu\nu} (p,q) \left( \frac{\alpha_p \beta_q}{p_0-q_0-k_0-i\epsilon}
- \frac{\alpha_q \beta_p}{p_0-q_0-k_0+i\eta} \right) } \nonumber\\
&& \quad \mbox{} +
T^{\mu\nu} (p,q') \left( \frac{\alpha_p \bar{\alpha}_p}{p_0+q_0-k_0-i\epsilon}
- \frac{\beta_p \bar{\beta}_q}{p_0+q_0-k_0+i\eta} \right) \nonumber\\
&& \quad \mbox{} +
T^{\mu\nu} (p',q) \left( \frac{\bar{\alpha}_p \alpha_q}{p_0+q_0+k_0-i\eta}
- \frac{\bar{\beta}_p \beta_q}{p_0+q_0+k_0+i\epsilon} \right) \nonumber\\
&& \quad \mbox{} +
T^{\mu\nu} (p',q') \left(
\frac{\bar{\alpha}_p \bar{\beta}_q}{p_0-q_0+k_0-i\epsilon}
- \frac{\bar{\beta}_p \bar{\alpha}_q}{p_0-q_0+k_0+i\eta} \right)
\biggr\rbrack
\label{2.12b}
\end{eqnarray}
\reseteqn
where
\begin{equation}
T^{\mu\nu} (p,q) \equiv \tr
\left( \gamma^\mu\ \gamma\cdot p\ \gamma^\nu\  \gamma\cdot q \right)
\label{2.13}
\end{equation}
\[
p'^{\mu} = (p^0, -\vec{p}), \qquad
q'^{\mu} = (q^0, -\vec{q}), \qquad
p^0 = |\vec{p}|, \qquad
q^0 = |\vec{q}| \qquad
\hbox{\rm and} \quad
\vec{p}=\vec{q}+\vec{k}.
\]
Further, $\eta=\epsilon$ for $T$ products, {\it i.e.\/} for
$\Pimnt$,  and $\eta=-\epsilon$ for the retarded function
$\Pimnr$.
Notice that the retarded function can  be obtained by continuing the real
part by the rule $k_0 \rightarrow k_0+i\epsilon$.
Formula \eqn{2.12b} agrees with Ref.~\cite{5} where time-ordered products
were used.  (We have changed the sign of $k$ in our definition  of the
Fourier transform relative to Ref.~\cite{5}.)

The real part of $\Pimn (k)$ is the same for both the time-ordered
and the retarded functions and has long been familiar \cite{1}.
Here we concentrate on the imaginary part
which is obtained from \eqn{2.12b} with the help of
\begin{equation}
\frac{1}{z-i\epsilon} = P \frac{1}{z} + i \pi \delta (z)
\label{2.14}
\end{equation}
Consider first the contribution due to the $T^{\mu\nu}(p,q')$
and $T^{\mu\nu}(p',q)$ terms in \eqn{2.12b}.
The imaginary parts carry the $\delta$-functions,
$\delta(p_0 + q_0 \pm k_0)$.  It is easily seen that these are
subdominant
at high temperature, for $k$ small compared to $T$.
[Their contributions are $O(T)$ or less.]
The dominant contributions to the imaginary part of $\Pimn$ come from
the $T^{\mu\nu}(p,q)$, $T^{\mu\nu}(p',q')$ terms.
For $\Pimnr$, we find from \eqn{2.12b}
\begin{eqnarray}
\Im \Pimnr =
\pi \int {\frac{d^3 q}{(2\pi)^3} \frac{1}{2p_0} \frac{1}{2q_0}}
&\hskip-.4em&
\biggl\lbrack T^{\mu\nu}(p,q) \, (n_q-n_p) \, \delta(p_0-q_0-k_0) \\
&\hskip-.4em&
\mbox{} -
T^{\mu\nu}(p',q') \, (\bar{n}_q-\bar{n}_p) \, \delta(p_0-q_0+k_0) \biggr\rbrack
\label{2.15}
\end{eqnarray}
For $\Pimnt$, we find
\begin{eqnarray}
\Im \Pimnt = \Im \Pimnr
+ 2\pi \int {\frac{d^3 q}{(2\pi)^3} \, \frac{1}{2p_0} \, \frac{1}{2q_0}}
&\hskip-.4em&
\biggl\lbrack T^{\mu\nu}(p,q) n_p(1-n_q) \delta(p_0-q_0-k_0) \\
&\hskip-.4em&  
\mbox{} +
T^{\mu\nu}(p',q') \bar{n}_q (1-\bar{n}_p) \delta(p_0-q_0+k_0) \biggr\rbrack
\nonumber
\label{2.16}
\end{eqnarray}
In the high temperature limit, $|\vec{k}|$ is small compared to $|\vec{p}|$,
$|\vec{q}|$, and there are simplifications \cite{5},
\begin{eqnarray}
T^{\mu\nu}(p,q) &\simeq& 8 q_0^2 Q^\mu Q^\nu \nonumber\\
T^{\mu\nu}(p',q') &\simeq& 8 q_0^2 Q'^{\mu} Q'^{\nu} \nonumber\\
p_0 - q_0 - k_0 &\simeq& -k \cdot Q \nonumber\\
p_0 - q_0 + k_0 &\simeq& k \cdot Q'
\label{2.17}
\end{eqnarray}
with $Q^\mu$, $Q'^{\mu}$ being the light-like four-vectors
\begin{equation}
Q^\mu = (1, \hat{q})\ , \qquad Q'^{\mu} = (1, -\hat{q})
\label{2.18}
\end{equation}

Equation \eqn{2.15} simplifies as
\begin{eqnarray}
\Im \Pimnr &\simeq& \frac{k_0 \Pmn}{2\pi^2}
\int_0^\infty {dq\ q(n_q+\bar{n}_q) } \nonumber\\
&=& \frac{k_0 T^2}{12} \Pmn
\label{2.19}
\end{eqnarray}
where
\begin{equation}
\Pmn = \int {d\Omega\ \delta (k \cdot Q) Q^\mu Q^\nu}
\label{2.20}
\end{equation}
and we take $n = \bar{n}$.  The integration in \eqn{2.20} is over the
orientations of the unit vector $\vec{Q}=\hat{q}$.
Note that from its definition, $\Pmn$ is transverse and traceless, while
the $\delta$-function enforces $k$ to be space-like.
Explicit evaluation gives
\begin{equation}
\Pmn = -k^2 \theta(-k^2) \frac{6\pi}{|\vec{k}|^3}
\left[ \frac{1}{3} \Pmn_1 + \frac{1}{2} \Pmn_2 \right]
\label{2.21}
\end{equation}
where
\alpheqn
\begin{equation}
\Pmn_1 = \left(g^{\mu\nu} - \frac{k^\mu k^\nu}{k^2} \right)
\label{2.22a}
\end{equation}
and
\begin{eqnarray}
P^{\mu 0}_2 &=& P^{0 \nu}_2 = 0 \nonumber\\
P^{ij}_2   &=& \delta^{ij} - \frac{k^i k^j}{|\vec{k}|^2}
\label{2.22b}
\end{eqnarray}
\reseteqn

Simplifying (2.16) similarly, we find
\begin{eqnarray}
\Im \Pimnt &\simeq&
\Im \Pimnr + \frac{\Pmn}{\pi^2}
\int_0^\infty {dq\,q^2 \, n_q (1-n_q) } \nonumber\\
&=& \Im \Pimnr + \frac{T^3}{6} \Pmn
\label{2.23}
\end{eqnarray}

This relationship between $\Pimnt$ and $\Pimnr$ can be understood in the
following way.  $\Pimnr$, being retarded,
obeys a spectral representation of the form \cite{2}
\begin{equation}
\Pimnr = \Pimnsub +
\int {d k'_0 \, \frac{\rmn (k_0', \vec{k})}{k_0'-k_0-i\epsilon}}
\label{2.24}
\end{equation}
for some spectral function $\rmn(k)$. $\Pimnsub$ is a `subtraction term'
that can arise in the real part of $\Pimnr$.
For $\Pimnt$, we then have \cite{2}
\begin{equation}
\Pimnt = \Pimnsub +
\int {d k_0'\,\frac{\rmn(k_0',\vec{k})}{k_0'-k_0-i\epsilon}
+ 2\pi i f(k_0) \rmn(k_0, \vec{k})}
\label{2.25}
\end{equation}
where
\begin{equation}
f(k_0) = \frac{1}{e^{k_0/T} - 1}
\label{2.26}
\end{equation}
The bosonic distribution function $f(k_0)$
appears because $\Pimnt$ is ultimately
part of the bosonic ({\it i.e.}~photon) propagator,
and also because it is given by
the thermal average of the $T$-product of two {\em bosonic} operators:
the two currents $j^\mu$ and $j^\nu$.
{}From \eqn{2.24} and \eqn{2.25}
we find
\begin{eqnarray}
\Im \Pimnr &=& \pi \rmn(k) \nonumber\\
\Im \Pimnt &=& \Im \Pimnr + 2\pi f(k_0) \rmn(k) \nonumber\\
&=& \pi \, \coth \, \frac{k_0}{2 T} \rmn(k)
\label{2.27}
\end{eqnarray}
The essence of our results \eqn{2.19} and \eqn{2.23} is that the
high-temperature spectral function is
\begin{equation}
\rmn(k) \simeq \frac{k_0 T^2}{12\pi} \Pmn
\label{2.28}
\end{equation}
and the difference in the high-temperature behavior between $\Im\Pimnr$ and
$\Im\Pimnt$ $[O(T^2)$ {\it vs.} $O(T^3)]$ is attributed to the
presence in the latter of
$2\pi f(k_0) \rmn$, which according to \eqn{2.26} and \eqn{2.28} tends to
$\frac{T^3}{6} \Pmn$.

The spectral function \eqn{2.28} also determines the high-temperature behavior
for the (common) real part of $\Pimnr$ and $\Pimnt$,
apart from possible subtraction terms.  Inserting \eqn{2.28} into the
dispersion formula \eqn{2.24} or \eqn{2.25} shows that the well-known high-T
asymptote for $\Pimn$ of \eqn{2.12b}
\begin{equation}
- \Re \Pimn \simeq \frac{T^2}{6}  \Pmn_2 +
\frac{T^2 k^2}{|\vec{k}|^2}
\left[ 1 + \frac{k_0}{2|\vec{k}|}
\ln \left| \frac{k_0 - |\vec{k}|}{k_0 + |\vec{k}|} \right| \right]
\left[ \frac{1}{3} \Pmn_1 + \frac{1}{2} \Pmn_2 \right]
\label{2.29}
\end{equation}
is reproduced with an appropriate subtraction term.

Our result \eqn{2.19}, \eqn{2.29}
for the retarded function
$\Pimnr$ agrees with various previous calculations \cite{1}, \cite{8}.
It is noteworthy that these early calculations in the Soviet literature, based
on the Boltzmann and Vlasov equations of kinetic theory, are here regained in
quantum field theory at one-loop order.

Another correlation function that is frequently considered
is the imaginary time one.  It too is given by
a dispersive integral \cite{2}.
\begin{eqnarray}
\Pimn_{\vbox{\offinterlineskip
\hbox{\the\scriptfont0 imaginary}
\vskip 1pt
\hbox{\the\scriptfont0 \quad time}}}
&=& \Pimnsub +
\int{ d k_0' \frac{\rmn (k_0', \vec{k})}{k_0' - \omega_n}} \nonumber\\
\omega_n &=& 2\pi i n T
\label{2.30}
\end{eqnarray}
Because the  ``external energy'' $\omega_n$ is
temperature dependent in imaginary time,
it makes sense to speak of high temperature behavior only for the $n=0$ mode,
(effectively reducing dimensionality to three)
where the spectral function enforces an $O(T^2)$ large-$T$ behavior.

The foregoing discussion emphasizes that finite-temperature field theory can
be described by {\em different} correlation functions, with {\em differing}
large-$T$ behavior.  In particular the retarded commutator is what is relevant
for the operator equations of motion, Landau damping, etc.  Because of the
asymmetry in its time-arguments,
the retarded commutator cannot be obtained by varying
an effective action, which necessarily involves the symmetric, time-ordered
products.  Correspondingly, only the retarded commutators
behave as $T^2$ for large $T$, while the time ordered products and the
real-time effective action have a more complicated large-$T$ behavior.
Finally, the imaginary time correlations again behave as $T^2$ for large $T$,
provided the ``external energies'' are restricted to the $n=0$ mode,
or are continued away from the temperature-dependent values
$2 \pi i n T$.
Such
analytic
continuation from the imaginary time expression
will produce the retarded functions in real time,
and not the time-ordered ones, which possess
non-analytic $[\sim f(k_0) \rmn]$ contributions.

In spite of this variety, a universally true statement can be made about the
high-temperature asymptote of the spectral function $\rmn$,
which determines quantities of physical interest
through the various dispersive representations.
In the present context, such a result is given in Eq.~\eqn{2.28}.

\section{Physical Significance of $\Im \Pimnr$}

The imaginary part of $\Pimnr$ describes Landau damping,
which can occur for fields with spacelike momenta \cite{8}.
Explicitly we consider
\alpheqn
\begin{equation}
\amx = \int {\frac{\dfour{k}}{(2\pi)^4}\,e^{-ikx} \amk}
\label{3.1a}
\end{equation}
with
\begin{equation}
\amk = \delta(k_0-\omega(\vec{k})) \amk + \delta(k_0+\omega(\vec{k})) \amdk
\label{3.1b}
\end{equation}
\reseteqn
and $\omega^2 < \vec{k}^2$.
The amplitudes $\amk$, $\amdk$, respectively of the positive and negative
frequency terms, correspond to absorption and emission processes.  The decay
of the field in the plasma arises from absorption by fermions and
antifermions.
The amplitude for absorption by fermions is given, to the lowest order in
coupling constant, by
\begin{equation}
{\cal A} = i \bar{u}_p (\gamma\cdot A) u_q (2\pi)^4
\delta^{(4)} (p-q-k) \sqrt{n_q (1-n_p)}
\label{3.2}
\end{equation}
where $u_p$, $u_q$ are wave functions for outgoing and incoming
fermions, respectively.
The factor $\sqrt{n_q (1-n_p)}$ arises
since the initial fermion is chosen from a state of occupation number $n_q$
and the final fermion is scattered into a state of occupation number $n_p$.
Single quantum absorption as in \eqn{3.2} is kinematically allowed for
spacelike momenta; it is in fact the inverse of \v{C}erenkov radiation.
One can also have creation of the mode $(\omega, \vec{k})$ by
(\v{C}erenkov) radiation from fermions, given by a formula like \eqn{3.2}
with $p\leftrightarrow q$ and $A\leftrightarrow A^\dagger$.
There are similar contributions from antifermions.
For the net absorption probability per unit spacetime volume,
denoted by $\gamma$, we then find, with summation over all fermion states,
\begin{eqnarray}
\gamma =
\int {\frac{d^3 q}{(2\pi)^3}\,\frac{1}{2p_0}\,\frac{1}{2q_0}}\,\amdk
&\hskip-.4em& 
\Biggl\lbrack T^{\mu\nu}(p,q) \left[ n_q (1-n_p) - n_p (1-n_q) \right]
2\pi\delta (p_0-q_0-k_0) \\
&\hskip-.4em&
\mbox{} +
T^{\mu\nu}(p',q')\left[\bar{n}_p(1-\bar{n}_q)-\bar{n}_q(1-\bar{n}_p)\right]
2\pi\delta (p_0-q_0+k_0) \Biggr\rbrack \ank \nonumber
\label{3.3}
\end{eqnarray}
where $p_0 = |\vec{p}|$, $q_0 = |\vec{q}|$.  Comparing this with \eqn{2.12b}
or \eqn{2.15} we see that
\begin{equation}
\gamma = 2 \amdk \left[ \Im \Pimnr(k_0) \right] \ank
\label{3.4}
\end{equation}
We can parametrize the field $A_\mu$ as
\begin{equation}
\vec{A} = \frac{\vec{k} k_0}{\vec{k}^2} A_0 + \vec{A}_T
\label{3.5}
\end{equation}
where $\vec{k}\cdot\vec{A}_T=0$.
$\vec{A}_T$ gives rise to the transverse electric field and the magnetic field;
$\phi = \left( 1-\frac{k_0^2}{\vec{k}^2} \right) A_0$
gives the longitudinal component of the electric field via
$\vec{E}_L = -\nabla \phi$.
In terms of the parametrization \eqn{3.5}
we can write
\begin{equation}
\gamma = T^2 \frac{\pi\omega}{|\vec{k}|}
\left[ \frac{1}{3} \phi^\dagger \phi
+ \frac{1}{6} \left(1-\frac{\omega^2}{\vec{k}^2} \right)
\vec{A}^{\dagger}_{T} \cdot \vec{A}_T \right]
\label{3.6}
\end{equation}
$\gamma$ is positive, as expected for net absorption or damping.

\section{Induced Current in a Non-Abelian Plasma}

We now consider the non-Abelian plasma and the induced current due to the hard
thermal loops.

As far as the two-point function is concerned, there is no significant
difference between the Abelian case of electrodynamics and QCD.
For the latter, the contribution of
$N_F$ flavors of quarks at high temperature is
\begin{equation}
\Im \Pi^{\mu\nu,ab}_{\,R} \simeq \delta^{ab} \frac{N_F}{2}
\left( \frac{k_0 T^2}{12} \Pmn \right)\ \ .
\label{4.1}
\end{equation}
The real part of the high-T two-point function is given by \cite{4}
\begin{equation}
\Re \Pi^{\mu\nu,ab} \simeq \delta^{ab} \frac{N_F T^2}{24\pi}
\left[ 4 \pi g^{\mu0} g^{\nu0}
- \int {d\Omega\,\frac{k_0 Q^\mu Q^\nu}{k \cdot Q}} \right]\ \ .
\label{4.2}
\end{equation}
The singularity when $k\cdot Q$ vanishes is defined as a principal value,
and then the integral reproduces \eqn{2.29}, apart from the group factors.

Notice that \eqn{4.1} and, with appropriate modifications,
\eqn{2.19} are obtained by continuing \eqn{4.2} by
$k_0 \rightarrow k_0 + i\epsilon$.
We expect this to be true in general, including the gluon contributions.
Thus we expect
\begin{equation}
\Im \Pi^{\mu\nu,ab}_{\,R} \simeq \delta^{ab} \left( N+\frac{N_F}{2} \right)
\frac{k_0 T^2}{12} \Pmn
\label{4.3}
\end{equation}
for $SU(N)$ gauge theory with $N_F$ flavors of fermions in the fundamental
representation.

We now consider the higher point functions in QCD.
The evolution of fields in the plasma is still described by an equation
similar to \eqn{2.2}, {\it viz.}~in the Feynman gauge
\begin{equation}
\Box A^\mu_a(x) = i S^{-1} \frac{\delta S}{\delta A_\mu^{a}(x)}
\label{4.4}
\end{equation}
This is still the operator equation of motion, reexpressed using the scattering
operator $S$.  The operator $S$ is now more complicated; correspondingly the
current defined by \eqn{4.4} is a more involved expression, including ghost
and gluon terms.
Nevertheless, when we expand \eqn{4.4} in powers of $A_\mu^a$,
we get only  retarded terms, since
$S^{-1}\frac{\delta S}{\delta A_\mu^a(x)}$
does not involve fields in the future
of $A_\mu^a(x)$.
In fact, writing
$U(x^0, y^0) = T \exp (i \int_{y^0}^{x^0} \dfourx {\cal L}_{\rm int}),
\ S=U(\infty,-\infty)$,
we find
\begin{equation}
J^\mu_a(x) = i S^{-1} \frac{\delta S}{\delta A_\mu^a (x)}
= U(-\infty, x^0)\, j^\mu_a (x) U(x^0, -\infty)
\label{4.5}
\end{equation}
where
$-\frac{\delta S_{\rm int}}{\delta A_\mu^a(x)} {\biggr\vert}_I = j^\mu_a(x)$.
When we differentiate \eqn{4.5} with respect to the potentials, we get
products of currents, and `contact' terms like
$\frac{\delta j^\mu_a(x)}{\delta A_\nu^b(y)}$.
Such terms eventually give
rise to expressions
like
$f^{a b c} A_\nu^b F_c^{\nu\mu}$
needed to write \eqn{4.4} with a gauge covariant derivative
$D_\nu F^{\nu\mu}$,
along with ghost and gauge-fixing terms.
[Strictly speaking, we need suitable $T^*$-products in the definition of $S$
to make the contact terms together with the commutator terms
come out covariant \cite{9}.
Our conclusion will not be affected by this needed qualification, since we use
only the retardation property.]\ \
Ignoring contact terms for the moment, we find
\begin{eqnarray}
\frac{\delta^n J^\mu_a(x)}
{\delta A^{b_1}_{\nu_1}(y_1) \cdots \delta A^{b_n}_{\nu_n}(y_n)}
= (-i)^n \sum_p
&\hskip-.4em&
\theta(x^0-y_1^0)
\theta(y_1^0-y_2^0)
\cdots
\theta(y_{n-1}^0-y_n^0) \nonumber\\
&\hskip-.4em& \quad \times\
[\ldots [ [ j_a^\mu(x),\,j_{b_1}^{\nu_1}(y_1)],\,j_{b_2}^{\nu_2}(y_2)] \ldots]
\label{4.6}
\end{eqnarray}
where the summation is over all permutations of
$(\nu_1\,b_1\,y_1,\ \nu_2\,b_2\,y_2,\ \ldots)$.
The right hand side of \eqn{4.6} is the definition of the multiple retarded
commutator.  Using this, we arrive at
\begin{eqnarray}
J_a^\mu(x) = j_a^\mu(x) +
\sum_{n=1}^\infty (-i)^n
&\hskip-.4em&
\int {
\dfour{y_1} \ldots \dfour{y_n}
\theta(x^0-y_1^0)
\theta(y_1^0-y_2^0)
\cdots
\theta(y_{n-1}^0-y_n^0) }
\nonumber\\
&\hskip-.4em&
\quad \times\
[\ldots[[j_a^\mu(x),\,j_{b_1}^{\nu_1}(y_1)],\,j_{b_2}^{\nu_2}(y_2)]\ldots]
A_{\nu_1}^{b_1}(y_1) \ldots A_{\nu_n}^{b_n}(y_n)
\label{4.7}
\end{eqnarray}
We must take thermal averages on the right hand side of \eqn{4.4}.
If we write
\begin{equation}
A_{\nu_1}^{b_1}(y_1) =
\int{\frac{\dfour{k_1}}{(2\pi)^4}\,e^{-i{k_1}{y_1}} A_{\nu_1}^{b_1}(y_1)},
\ {\rm etc.,}
\label{4.8}
\end{equation}
the retardation property of \eqn{4.7} indicates that the integration over the
time-components of $y_i$'s requires
$k_i^0 \rightarrow k_i^0 + i\epsilon$
for convergence.
This will generally be true,
since it follows from the retardation property and does not depend
specifically on the expansion \eqn{4.7}.
We can thus make a rule for the $n$-point functions:
Calculate the thermal average of the current, first ignoring the
$i\epsilon$'s in the propagators or energy-denominators.  Then for each
momentum $k_i$ at each of the $y_i$'s (but not at the unintegrated point $x$),
make the replacement $k_i^0 \rightarrow k_i^0 + i\epsilon$.  The averaged
equation of motion \eqn{4.4}
\begin{equation}
\Box A_\mu^a(x) = \langle J_\mu^a (x) \rangle
\label{4.9}
\end{equation}
gives the evolution of fields in the plasma,
$\langle J_\mu^a(x) \rangle$ being calculated as above.

We can now apply this argument to the induced current due to hard thermal
loops in QCD.  The hard thermal loop contribution,
found in Ref.~\cite{5},
is determined in the following fashion.
We begin with the functional $I(A)$ defined on Euclidean 2-space
[complex coordinates $z$ and $\bar{z}$] and depending on the Lie-algebra
valued vector potential $A$, with a single (spatial) component \cite{10}.
\begin{eqnarray}
I(A) = i \sum^\infty_{n=2} \frac{(-1)^n}{n}
\int {\frac{d^2{z_1}}{\pi} \cdots \frac{d^2{z_n}}{\pi}
\frac{\tr A(z_1,\bar{z}_1) \ldots A(z_n,\bar{z}_n)}
{\bar{z}_{1 2} \bar{z}_{2 3}  \ldots \bar{z}_{n-1\,n} \bar{z}_{n 1}}}
\label{4.10}
\end{eqnarray}
\[ \bar{z}_{i j} \equiv \bar{z}_{i} - \bar{z}_{j}\ \ . \]
This is the Chern-Simons eikonal,
{\it i.e.\/}~it is the exponent in a WKB wave functional.
In a Chern-Simons gauge theory on (2+1) dimensional space-time,
the Schr\"odinger-picture state, which is defined on 2-space (at fixed time)
and satisfies the Chern-Simons Gauss law,
has the form $\bar{e}^{I(A)}$,
with $A$ being one of the two spatial gauge potentials of the theory
and the other being conjugate to $A$.
Owing to the simple symplectic structure of this theory,
the eikonal/WKB approximation is exact, and $I(A)$
coincides with the eikonal.
$I(A)$
is also recognized as the Polyakov-Wiegman determinant
for two-dimensional, single chirality fermions \cite{6}.
The structure \eqn{4.10} is relevant to
high-T field theory owing to the fact that
its gauge transformation properties are closely related to those of the
generating functional for hard thermal loops \cite{5},
and the latter is completely determined by its response to gauge
transformations.

To present the thermal
generating functional \eqn{4.10} must be elaborated upon and
continued from the Euclidean space $(z, \bar{z})$ to Minkowski space-time.
This is accomplished by replacing $A$ in \eqn{4.10} with
$A_{+} \equiv \frac{1}{2} A \cdot Q$, while
$z$ is replaced by $x \cdot Q'$,
$\bar{z}$ by $x \cdot Q$.
Also all the $A$'s $\rightarrow \frac{1}{2}A\cdot Q$
depend on four variables:
$z \rightarrow x \cdot Q'$,
$\bar{z} \rightarrow x \cdot Q$
and
$\vec{x}_\bot$, which is orthogonal to
$Q$ and $Q'$.
All the $A$'s carry the same value of $\vec{x}_\bot$,
which therefore is a fixed parameter in the functional \eqn{4.10},
and is subjected to a single integration:
$I(A) \rightarrow \int {d^2 \vec{x}_\bot I(A) \equiv {\cal I}(A)}$
The hard thermal loop generating functional is now given by
\begin{equation}
\Gamma = - \left( N + \frac{N_F}{2} \right) \frac{T^2}{6\pi}
\int {d\Omega \left[
\int {\dfourx \tr A_{+} A_{-} + i\pi {\cal I}(A_{+})
+ i\pi {\cal I}' (A_{-})} \right] }
\label{4.11}
\end{equation}
where ${\cal I}'(A_{-})$ is obtained from ${\cal I}(A_{+})$
by interchanging
$Q$ and $Q'$, and $A_{-} \equiv \frac{1}{2} A \cdot Q'$.

The integrals in Minkowski space-time are now singular: when
transformed to momentum space they contain denominators
involving $k\cdot Q$ and $k\cdot Q'$, which can vanish.
If these zeroes are ignored and the integrals are evaluated formally, one
regains the hard thermal loop contributions, but they do not have the
appropriate analyticity properties, and cannot be interpreted as time-ordered
products, retarded products, etc.

Here we give the prescription which results in an evaluation of
the retarded current, relevant to the non-Abelian gauge theory.

The contribution of ${\cal I}(A_{+})$ in \eqn{4.11} to the current is
given in Euclidean space by \cite{10}
\begin{equation}
-\frac{\delta{\cal I}(\frac{1}{2} A \cdot Q)}{\delta A_\mu^a(x)}
=
- \left( N + \frac{N_F}{2} \right) \frac{T^2}{6\pi}
\int
{
{d\Omega \sum_{n=1}^{\infty} (-1)^{n+1}
\int {\frac{d^2{z_1}}{\pi} \cdots \frac{d^2{z_n}}{\pi}}}
\ \frac{\tr \left[ \left( \frac{T_a Q^\mu}{2} \right)
A_{+}(x_1) \cdots A_{+}(x_n) \right]}
{(\bar{z} - \bar{z}_1) \bar{z}_{1 2} \cdots \bar{z}_{n - 1n}
(\bar{z}_n - \bar{z})}
}
\label{4.12}
\end{equation}
where $T_a$ is the anti-Hermitian group generator.
Using \eqn{4.8} and
\begin{equation}
\frac{1}{\bar{z}} = \frac{2\pi}{i} \int
{\frac{d^2 p}{(2\pi)^2}\, \frac{e^{ipx}}{\bar{p}}}
\label{4.13}
\end{equation}
we can write \eqn{4.12} in [Euclidean] momentum space.
The term with $n$ potentials is given by
\alpheqn
\begin{equation}
-\frac{\delta {\cal I}_n (k)}{\delta A_\mu^a} =
\left( N + \frac{N_F}{2} \right) \, \frac{T^2}{6\pi} \, \frac{(2i)^{n+1}}{4}
\tr\left[\left(\frac{T_a Q^\mu}{2}\right) A_{+}(k_1) \ldots A_{+}(k_n) \right]
F (k_1, \ldots, k_n)
\label{4.14a}
\end{equation}
where
\begin{equation}
F = -4\pi \int { \frac{d^2p}{(2\pi)^2}\,\frac{1}
{(\bar{p}+\bar{q}_0)  (\bar{p}+\bar{q}_1) \ldots (\bar{p}+\bar{q}_n)}}
\label{4.14b}
\end{equation}
\reseteqn
\begin{equation}
\bar{q}_0 = 0 \ \ , \hskip.5in \bar{q}_i = \sum_{j=1}^i \bar{k}_j
\label{4.15}
\end{equation}
Using the identities
\alpheqn
\begin{equation}
\int{\frac{d^2 p}{(2\pi)^2}\,\frac{\partial}{\partial p}
\left[ p \Pi_i \frac{1}{(\bar{p}+\bar{q}_i)} \right] = 0 }
\label{4.16a}
\end{equation}
\begin{equation}
\frac{\partial}{\partial p}  \left( \frac{1}{\bar{p}+\bar{q}} \right) =
\pi \delta^{(2)} (p + q)
\label{4.16b}
\end{equation}
\reseteqn
we can evaluate $F$ as
\begin{equation}
F(k_1, k_2, \ldots, k_n) = \sum_{i=0}^n
\frac{- q_i}
{ (\bar{q}_0 - \bar{q}_i)(\bar{q}_1 - \bar{q}_i)
\cdots
(\bar{q}_{i-1} - \bar{q}_i)(\bar{q}_{i+1} - \bar{q}_i)
\cdots
(\bar{q}_n - \bar{q}_i)}
\label{4.17}
\end{equation}
We can now continue to Minkowski space by
$\bar{k} \rightarrow k \cdot Q$,
$k \rightarrow k \cdot Q'$.
Since the $A$'s have $e^{-i k x}$ factors,
the retardation condition requires
$k_i^0 \rightarrow k_i^0 + i \epsilon_i$.
We thus define $F$ in Minkowski space as the expression \eqn{4.17}
with
\begin{eqnarray}
\bar{q}_i &=& \sum_{j=1}^i (k_j \cdot Q + i \epsilon_j)\ , \nonumber\\
q_i &=& \sum_{j=1}^i k_j \cdot Q'\ .
\label{4.18}
\end{eqnarray}
The expression for the current, which follows, is therefore written as
\cite{11}
\alpheqn
\begin{equation}
J_a^\mu(x) = \sum_{n=1}^\infty
\int \frac{\dfour{k_1}}{(2\pi)^4} \cdots \frac{\dfour{k_n}}{(2\pi)^4}\,
e^{-i (\sum\limits^n k) \cdot x} J_{a,n}^\mu (k)
\label{4.19a}
\end{equation}
\begin{eqnarray}
J_{a,n}^\mu (k) &=&
\left( N + \frac{N_F}{2} \right) \frac{T^2}{6\pi}
\int {
d\Omega \, \Biggl\lbrack
\tr \left\{ \left( \frac{T_a Q^\mu}{2} \right)
A_{-}(k_1) + A_{+}(k_1) \left( \frac{T_a Q^\mu}{2} \right) \right\}
\delta_{n,1}
} \\
&& \qquad \mbox{} + \biggl\lbrace
\frac{(2i)^{n+1}}{4} \tr
\left[ \left( \frac{T_a Q^\mu}{2} \right)
A_{+}(k_1) \ldots  A_{+}(k_n) \right]
F (k_1, \ldots, k_n)
+\ \left( Q \leftrightarrow Q' \right) \biggr\rbrace \Biggr\rbrack \nonumber
\label{4.19b}
\end{eqnarray}
\reseteqn
The equations of motion for the potential with the hard thermal loop
corrections is
\begin{equation}
\left( D_\nu F^{\nu\mu} \right)_a = J_a^\mu
\label{4.20}
\end{equation}
with $J_a^\mu$ from (4.19).
(Evidently we have not displayed gauge-fixing and ghost terms.)
\ \ Equation \eqn{4.20} describes Landau damping of fields in the quark-gluon
plasma as well as Debye screening and propagation of plasma waves in a gauge
covariant way.
[Of course, as mentioned in the Introduction, these equations still do not
include the soft momentum loop-integrations which have to be done in terms of
effective vertices and propagators defined by equations
(4.19), \eqn{4.20}.]

The two-point function ({\it i.e.\/}~$n=1$)
as defined by (4.19)
agrees with our previous considerations.  The three-point function
({\it i.e.\/}~$n=2$) as given again by (4.19) is
\begin{eqnarray}
J_{a,2}^{\mu} (k) = 2 i
\left( N + \frac{N_F}{2} \right) \frac{T^2}{6\pi}
&\hskip-.4em&
\int { d\Omega\ \Biggl\lbrace
\tr \left[ \left( \frac{T_a Q^\mu}{2} \right)
A_{+}(k_2) A_{+}(k_3) \right] } \\
&\hskip-.4em& \quad
\times\
\frac
{\left(
k_2 \cdot Q \,
k_3 \cdot Q'
- k_2 \cdot Q' \,
k_3 \cdot Q \right)}
{(k_2 \cdot Q + i \epsilon)
(k_3 \cdot Q + i \epsilon)
\left[ (k_2 + k_3) \cdot Q + i \epsilon \right]}
+ \left( Q \leftrightarrow Q' \right) \Biggl\rbrace \nonumber
\label{4.21}
\end{eqnarray}
For quark loops we do not have to consider $T^{*}$-products or
contact terms like
$\frac{\delta j^\mu_a}{\delta A_\nu^b}$ since
$j^{\mu,a} = i \bar{q}_I \gamma^\mu T^a q_I$.
Equation \eqn{4.8} holds as it is written and we can calculate the 3-point
function (i.e.~two $A$'s) by evaluating the retarded commutators.
One can verify that this does indeed give equation (4.21) with the
$i \epsilon$'s as indicated.

Our emphasis has been on
the operator equations of motion and on how retarded functions
arise in such a description.  There are of course many other real time
formalisms for thermal field theory.  For example, in the Keldysh approach
\cite{8} one considers time to run along a contour,
which passes from $-\infty$ to $\infty$,
and then folds back and returns from $\infty$ to $-\infty$.
We have checked that the evolution equations for field configurations in this
formalism lead to the retarded functions as in our discussion.

Prescriptions
for obtaining the imaginary parts of Green's functions in terms of
`cutting rules' have been discussed in Ref.~\cite{12}.

\section*{Acknowledgement}
We have benefitted from conversations with
L.~Madansky,
R.~Pisarski, A.~Rebhan and H.~Weldon;
these we gratefully acknowledge.


\begin{thebibliography}{99}
\bibitem{1}
V.~P.~Silin,
{\it Zh.~Eksp.~Teor.~Fiz.\/} {\bf 38}, 1577 (1960)
[Engl.~trans.: {\it Sov.~Phys.~JETP\/} {\bf 11}, 1136 (1960)];
V.~N.~Tsytovich,
{\it Zh.~Eksp.~Teor.~Fiz.\/} {\bf 40}, 1775 (1961)
[Engl.~trans.: {\it Sov.~Phys.~JETP\/} {\bf 13}, 1249 (1961)];
O.~K.~Kalashnikov and V.~V.~Klimov,
{\it Yad.~Fiz.\/} {\bf 31}, 1357 (1980)
[Engl.~trans.: {\it Sov.~J.~Nucl.~Phys.\/} {\bf 31}, 699 (1980)];
V.~V.~Klimov,
{\it Yad.~Fiz.\/} {\bf 33}, 1734 (1981)
[Engl.~trans.: {\it Sov.~J.~Nucl.~Phys.\/} {\bf 33}, 934 (1981)],
{\it Zh.~Eksp.~Teor.~Fiz.\/} {\bf 82}, 336 (1982)
[Engl.~trans.: {\it Sov.~Phys.~JETP\/} {\bf 55}, 199 (1982)]
\bibitem{2}
L.~Dolan and R.~Jackiw,
{\it Phys.~Rev.~D\/} {\bf 9}, 3320 (1974)
\bibitem{3}
R.~Pisarski,
{\it Physica\/} {\bf A158}, 246 (1989);
{\it Phys.~Rev.~Lett.\/} {\bf 63}, 1129 (1989).
\bibitem{4}
E.~Braaten and R.~Pisarski,
{\it Phys.~Rev.~D\/} {\bf 42}, 2156 (1990);
{\it Nucl.~Phys.\/} {\bf B337}, 569 (1990);
{\it ibid.\/} {\bf B339}, 310 (1990);
{\it Phys.~Rev.~D\/} {\bf 45}, 1827 (1992);
J.~Frenkel and J.~C.~Taylor,
{\it Nucl.~Phys.\/} {\bf B334}, 199 (1990);
J.~C.~Taylor and S.~M.~H.~Wong,
{\it Nucl.~Phys.\/} {\bf B346}, 115 (1990).
\bibitem{5}
R.~Efraty and V.~P.~Nair,
{\it Phys.~Rev.~Lett.\/} {\bf 68}, 2891 (1992);
Columbia Preprint CU--TP 579, Nov.~1992
(to be published in {\it Phys.~Rev.~D.\/})
\bibitem{6}
D.~Gonzalez and A.~Redlich,
{\it Ann.~Phys.\/} (NY) {\bf 169}, 104 (1986);
G.~Dunne, R.~Jackiw and C.~Trugenberger,
{\it Ann.~Phys.\/} (NY) {\bf 149}, 197 (1989).
\bibitem{7}
R.~Kubo,
{\it J.~Phys.~Soc.~Japan\/} {\bf 12}, 570 (1957).
\bibitem{8}
E.~M.~Lifshitz and L.~P.~Pitaevskii,
{\it Physical Kinetics}
(Pergamon, Oxford, 1981).
\bibitem{9}
D.~Gross and R.~Jackiw,
{\it Nucl.~Phys.\/} {\bf B14}, 269 (1969)
\bibitem{10}
The first term in the series
involves a product of distributions
$-\frac{1}{(\bar{z}_{1 2})^2}$;
this is defined as
$\frac{\partial}{\partial \bar{z}_1} \, \frac{1}{\bar{z}_{1 2}}$.
\bibitem{11}
A summed formula for the current in Euclidean space can be given, see
Ref.~\cite{6}.
\[ J_{+} \propto A_{+} - g^{-1} \partial_{+} g \]
\[ J_{-} \propto A_{-} - h^{-1} \partial_{-} h \]
Here
$A_{+} = h^{-1} \, \partial_{+} \, h$,
$A_{-} = g^{-1} \, \partial_{-} \, g$
and $\pm$ refers to the $(1)\pm i(2)$ components.
However, no compact expression in Minkowski space encodes the proper
$i\epsilon$ prescription.
\bibitem{12}
H.~A.~Weldon,
{\it Phys.~Rev.~D\/} {\bf 28}, 2007 (1983);
R.~L.~Kobes and G.~W.~Semenoff,
{\it Nucl.~Phys.\/} {\bf B260}, 714 (1985).
\end{thebibliography}
\end{document}